\documentclass[11pt]{article}
\usepackage{graphicx}
\usepackage{amssymb}
\usepackage{amsmath}
\usepackage{graphicx}
\usepackage{amstext}
\usepackage{esint}
\usepackage{inputenc}

\def\beq{\begin{eqnarray}}    %%%  begequation/eqnarray
\def\eeq{\end{eqnarray}}      %%%  endequation/eqnarray
\newcommand{\be}{\begin{equation}}
\newcommand{\ee}{\end{equation}}
\newcommand{\rma}{\rho_m}
\newcommand{\rr}{\rho_r}
\newcommand{\rL}{\rho_\Lambda}

\newcommand{\CC}{\Lambda}

\newcommand{\rD}{\rho_D}

\newcommand{\wD}{\omega_D}

\newcommand{\nueff}{\nu_{\rm eff}}

\begin{document}
%\pubblock

%\today

%\vspace{1cm}

 \hyphenation{nu-cleo-syn-the-sis u-sing si-mu-la-te ma-king
cos-mo-lo-gy know-led-ge e-vi-den-ce stu-dies be-ha-vi-or
res-pec-ti-ve-ly appro-xi-ma-te-ly gra-vi-ty sca-ling
ge-ne-ra-li-zed}

%%%%%%%%%%%%%%%%%%%%%%%%%%%%%%%%%%%%%%%%%%%%%%%%%%%%%%%%%

%\newpage

%%%%%%%%%%%%%%%%%%%%%%%%%%%%%%%%%%%%%%%%%%%%%%%%%%%%%%%%%
%\flushright{UB-ECM-PF-06/24 }\\

\begin{center}
{\it\Large Running Vacuum in the Universe: current phenomenological status\footnote{Invited talk at the 14th Marcel Grossmann Meeting - MG14,  Univ. of Rome ``La Sapienza'' - Rome, Italy, 2015}}

\vskip 2mm

 \vskip 8mm

\textbf{Joan Sol\`{a}}

\vskip 0.5cm

$^1$High Energy Physics Group, Departament ECM, Universitat de Barcelona,\\
Av. Diagonal 647, E-08028 Barcelona, Catalonia, Spain

\vskip0.15cm

and

\vskip0.15cm

Institute of Cosmos Sciences,
Universitat de Barcelona

\vskip0.4cm

E-mail:  sola@ecm.ub.edu

 \vskip2mm

\end{center}
\vskip 15mm

\begin{quotation}
\noindent {\large\it \underline{Abstract}}. I review the excellent phenomenological status of a class of dynamical vacuum models in which the vacuum energy density, $\rL=\rL(H)$, as a function of the Hubble rate, evolves through its interaction with dark matter and/or through the accompanying running of the gravitational coupling $G$, including the possibility of being self-conserved with a nontrivial effective equation of state. Some of these models have been used to incorporate into a single vacuum structure the rapid stage of inflation, followed by the standard radiation and cold dark matter epochs all the way down until the dark energy era. Remarkably, the running vacuum models (RVM's) render an outstanding phenomenological description of the main cosmological data at a level that is currently challenging the concordance $\Lambda$CDM model, thereby implying that present observations seem to point to a running vacuum rather than to a rigid cosmological constant $\Lambda$ in our Universe.
\end{quotation}
\vskip 5mm

\newpage

%\tableofcontents

\newpage

%%%%%%%%%%%%%%%%%%%%%%%%%%%%%%%%%%%%%%%%%%%%%%%%%%%%%%%%%%%%%%%%%
%%%%%%%%%%%%%%%%%%%%%%%%%%%%%%%%%%%%%%%%%%%%%%%%%%%%%%%%%%%%%%%%%
%%%%%%%%%%%%%%%%%%%%%%%%%%%%%%%%%%%%%%%%%%%%%%%%%%%%%%%%%%%%%%%%%

\section{Introduction}

The cosmological term in Einstein's equations is approaching a century of existence\,\cite{Einstein1917}. It has traditionally been associated to the concept of vacuum energy density: $\rL = \CC/(8\pi G)$. Only after the advent of the quantum theory and quantum field theory (QFT) this connection acquired some meaning. However, it also became more troublesome since it triggered the famous (so far unsolved) cosmological constant (CC) problem\,\cite{CCproblem1}. This problem is the main source of headache for every theoretical cosmologist confronting his/her theories with the measured value of $\rL$\,\cite{PLANCK2015}. Furthermore, the purported discovery of the Higgs boson at the LHC has accentuated the CC problem greatly, certainly much more than is usually recognized\,\cite{JSPReview2013,SolaGomez2015}.

Owing to the necessary spontaneous symmetry breaking (SSB) of the electroweak (EW) theory, an induced contribution  to $\rL$ is generated which is appallingly much larger (viz. $\sim 10^{56}$) than the tiny value $\rL\sim 10^{-47}$ GeV$^4$ (``tiny'' only within the particle physics standards, of course) extracted from observations. So the world-wide celebrated  ``success'' of the Higgs finding  in particle physics actually became a cosmological fiasco, since it instantly detonated the ``modern CC problem'', i.e. the confirmed size of the EW vacuum, which should be deemed  as literally ``real'' (in contrast to other alleged -- ultralarge -- contributions from QFT)  or ``unreal'' as the Higgs boson itself! One cannot exist without the other. Does this mean that the found Higgs boson is not a fundamental particle? Think seriously about it!

I refer the reader to some review papers\,\cite{CCproblem1}, including \cite{JSPReview2013,SolaGomez2015}, for a more detailed presentation of the CC problem. %and its connection to vacuum energy.
Setting aside the ``impossible'' task of predicting the $\CC$ value itself -- unless it is understood as a ``primordial renormalization''\,\cite{GRF2015} -- I will focus here on a special class of models in which $\CC$ appears neither as a rigid constant nor as a scalar field (quintessence and the like)\,\cite{CCproblem1}, but as a ``running'' quantity in QFT in curved spacetime. This is a natural option for an expanding Universe. As we will show, such kind of dynamical vacuum models are phenomenologically quite successful; in fact so successful that they are currently challenging the $\CC$CDM\,\cite{ApJ2015,JCAP2015b,MNRAS2015,JCAP2015a,BPS2009}.

%\subsection{Subsections only have the first letter of the entire title capitalized}

%Subsections only have the first letter of the first word capitalized (except for words that are naturally capitalized).

\section{Running Vacuum Models}

The running vacuum models (RVM's)\,(cf. \cite{JSPReview2013,SolaGomez2015} and references therein) are based on the idea that the cosmological term $\CC$, and the corresponding vacuum energy density, $\rL$,  should be time dependent quantities in cosmology. It is difficult to conceive an expanding universe with a strictly constant vacuum energy density that has remained
immutable since the origin of time. Rather, a smoothly evolving DE density that inherits its time-dependence
from cosmological variables $\mu=\mu(t)$, such as the Hubble rate
$H(t)$ or the scale factor $a(t)$, is not only a qualitatively more
plausible and intuitive idea, but is also suggested by fundamental
physics, in particular by QFT in curved
space-time. We denote it in general by $\rD=\rD(\mu(t))$, as it may have an effective equation of state (EoS) $\wD=\wD(H)$  more general than that of the vacuum ($\wD=-1$).  The main standpoint of the RVM class of dynamical DE models is that $\rD$ ``runs'' because the effective action receives quantum effects from the matter fields. The leading effects may generically be captured from a renormalization group equation (RGE) of the form\,\cite{JSPReview2013}
\begin{eqnarray}\label{seriesRLH}
\frac{d\rD}{d\ln
\mu^2}=\frac{1}{(4\pi)^2}\sum_{i}\left[\,a_{i}M_{i}^{2}\,\mu^{2}
+\,b_{i}\,\mu^{4}+c_{i}\frac{\mu^{6}}{M_{i}^{2}}\,+...\right] \,.
\end{eqnarray}
The RVM ansatz is that $\rD=\rD(H)$  because $\mu$ will be naturally associated to the Hubble parameter at a given epoch $H=H(t)$, and hence $\rD$ should evolve with the rate of expansion $H$. Notice  that $\rD(H)$ can involve \emph{only} even powers of
the Hubble rate $H$ (because of the covariance of the effective
action)\,\cite{JSPReview2013}. The coefficients $a_i$,
$b_i$,$c_i$... are dimensionless, and the $M_i$ are the masses of the particles in the loops. Because $\mu^2$ can be in general a linear combination of the homogeneous terms $H^2$ and $\dot{H}$, it is obvious that upon integration of the above RGE we expect the following general type of (appropriately normalized) RVM density\,\cite{JSPReview2013,SolaGomez2015,GRF2015}:
\begin{eqnarray}\label{eq:ModelsA}
\rho_D(H)&=&\frac{3}{8\pi
G}\left(C_0+\nu H^2+\frac{2}{3}\alpha \dot{H}\right)+{\cal O}(H^4)\,,
\label{eq:ModelsC}
\end{eqnarray}
We emphasize that $C_0\neq 0$ so as to insure a smooth $\CC$CDM limit when the dimensionless coefficients $\nu$ and $\alpha$ are set to zero\,\footnote{It is important to make clear that models with $C_0=0$ (for any $\nu$ and $\alpha$) are ruled out by the observations, as shown in \cite{JCAP2015b,MNRAS2015,JCAP2015a}. This conclusion also applies to all DE models of the form  $\rD\sim aH+bH^2$,
with a linear term $\sim H$ admitted only on phenomenological grounds\,\cite{JCAP2015b,MNRAS2015}. In particular, the model $\rD\sim H$ is strongly ruled out, see \cite{MNRAS2015} (and the discussion in its Appendix).}. The interesting possibility that $\nu$ and/or $\alpha$ are nonvanishing may induce a time evolution of the vacuum energy. These dimensionless coefficients can be computed in QFT from the ratios squared of the masses to the Planck mass\,\cite{Fossil07}, and are therefore small as expected from their interpretation as $\beta$-function coefficients of the RGE (\ref{seriesRLH}). Since some of the masses inhabit the GUT scale $M_X\sim 10^{16}$ GeV, the values of $\nu,\alpha$ need not be very small, typically $\sim 10^{-3}$ at most upon accounting for the large multiplicities that are typical in a GUT -- see Ref.\cite{Fossil07} for a concrete estimate. Ultimately, $\nu$ and $\alpha$ must be determined phenomenologically by confronting the model to the wealth of observations. It is remarkable that the aforementioned theoretical estimate is of the order of magnitude of the phenomenological determination\,\cite{ApJ2015,JCAP2015b,MNRAS2015,JCAP2015a,BPS2009}. Finally, we note that the ${\cal O}(H^4)$-terms in (\ref{eq:ModelsA}) are irrelevant for the study of the current Universe, but are essential for the correct account of the inflationary epoch in this context. The RVM (\ref{eq:ModelsA}) is therefore capable of providing a unified dynamical vacuum picture for the entire cosmic evolution\,\cite{BLS,BMS}. It also has the ability to explain the graceful exit and entropy problems\,\cite{GRF2015}. Let us also mention that the relation of the RVM with entropic and QCD-ghost models is discussed in \cite{JCAP2015b,BasPolSol2012,KomatsuKimura}(cf. also the previous footnote).

From the explicit expression (\ref{eq:ModelsA}) of the RVM one can solve for the cosmological equations relevant in the current Universe:
 \begin{equation}\label{eq:generalizedFriedmann}
3H^2=8\pi
G(\rho_m+\rD(H))\,,\ \ \ \ \ \ \ \  2\dot{H}+3H^2=-8\pi
G(\omega_m\rho_m+\omega_D\rD(H))\,,
\end{equation}
where  $\omega_m$ and $\omega_D$ are the  EoS parameters of the matter fluid and of the DE, respectively. The explicit solution will depend of course on whether we assume that the DE is canonical vacuum energy ($\omega_D=-1$), in which case $\rD$ can be properly denoted as $\rL$, or dynamical DE with a nontrivial EoS evolving with time, $\omega_D=\omega_D(t)$ (with $\omega_D(t_0)\simeq -1$ now). It will also depend on whether the gravitational coupling $G$ is constant or also running  with the expansion, $G=G(H)$ (as $\rD$ itself). And, finally, it will depend on whether we assume that there exists an interaction of the DE with the matter (mainly dark matter, DM). Whatever it be the  nature of our assumptions on these important details, they must be of course consistent with the Bianchi identity, which is tantamount to say with the local covariant conservation laws. In fact, these possibilities have all been carefully studied in the literature and the complete solution of the cosmological equations have been provided in each case. We refrain of course from writing out the details in this short review; the reader can find them in full in the comprehensive studies\,\cite{ApJ2015,JCAP2015b,MNRAS2015,JCAP2015a,BPS2009}.  Let us, however, show the solutions for the densities  of matter and DE in the case $\omega_D=-1$ and assuming there is interaction between the two media at fixed $G$. The local conservation law of matter and vacuum then reads $\dot{\rho}_{\CC}+\dot{\rho}_m+3H(\rho_m+p_m)=0$. In terms of the cosmological redshift $z$ one finds:
\begin{equation}\label{MatterdensityCCtCDM}
\rho_m(z) = \rho_m^0 ~(1+z)^{3 \xi}+\rho_r^0 (1+z)^{4 \xi'}\,,
\end{equation}
where $\rho_m^0$ and $\rho_r^0$ are the current values of cold matter and radiation. Similarly, the dynamical DE density  (of vacuum type, in this case) reads
\begin{equation}\label{CCdensityCCtCDM}
\rL(z)=\rL^0+{\rho_m^0}\,\,(\xi^{-1} - 1) \left[(1+z)^{3\xi} -1  \right]+
{\rho_r^0}\,\,(\xi'^{-1} - 1) \left[ (1+z)^{4\xi'} -1\right]\,,
\end{equation}
with
\begin{equation}\label{defxiM}
\xi= \frac{ 1 - \nu }{ 1 - \alpha }\,, \ \ \ \ \ \ \ \ \ \xi'= \frac{ 1 - \nu }{ 1 - 4\alpha/3 }\,.
\end{equation}
\begin{figure}[t]
\begin{center}
\includegraphics[width=6in]{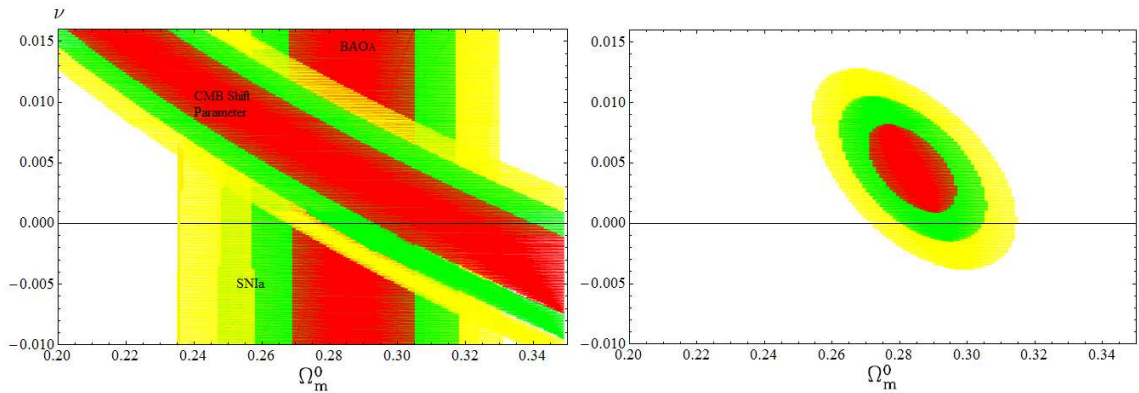}
\end{center}
\caption{Left: Intersection of the SNIa$+$BAO$+$ CMB shift parameter data in the $(\Omega_m^0, \nu)$ plane for the running vacuum model (\ref{eq:ModelsA}) with $\alpha=0$, assuming $G=$const. and allowing interaction between vacuum ($\omega_D=-1$) and DM. Right: The resulting 1, 2, 3$\sigma$ likelihood contours. The $\CC$CDM case is the $\nu=0$ line.}
\label{aba:fig1}
\end{figure}

As we can see from (\ref{CCdensityCCtCDM}), the vacuum behaves as a pure CC term $\rL=\rL^0$ only if $\xi=\xi'=1$. From (\ref{defxiM}) we observe that we can remain close to it if  $|\nu|, |\alpha|\ll 1$. At the same time, $\xi\neq 1$ (and/or $\xi'\neq 1$) implies an anomalous conservation law for matter-radiation, see (\ref{MatterdensityCCtCDM}). Obviously we can afford this situation only if the mentioned tiny deviations from the standard conservation laws are allowed. But even for values as small as  $|\nu|, |\alpha|\lesssim 10^{-3}$ is sufficient to infer observational evidence of physics beyond the $\CC$CDM, as we will see in the next section.

\section{Fitting the observational data to the Running Vacuum}

In this section we present a summary of the numerical results obtained after comparing the general class of RVM's (\ref{eq:ModelsA}) with all the main sources of cosmological data collected up to date, which include the expansion history data on distant supernovae (SNIa), the Baryonic Acoustic Oscillations (BAO), the data on the Hubble function $H(z_i)$ at different redshifts; and, of course, also the cosmic microwave background (CMB) observations and the linear growth data on the formation of large scale structures (LSS)\footnote{The RVM model was first tested against SNIa data in \cite{Cristina2003}. More recently, it was analyzed in great detail with the inclusion of all the mentioned sources of data in \cite{ApJ2015,JCAP2015b,MNRAS2015,JCAP2015a,BPS2009}. In these references one can also find the entire list of observational works from where all the data used in our analysis have been extracted.}.
\begin{figure}[t]
\begin{center}
\includegraphics[width=6in]{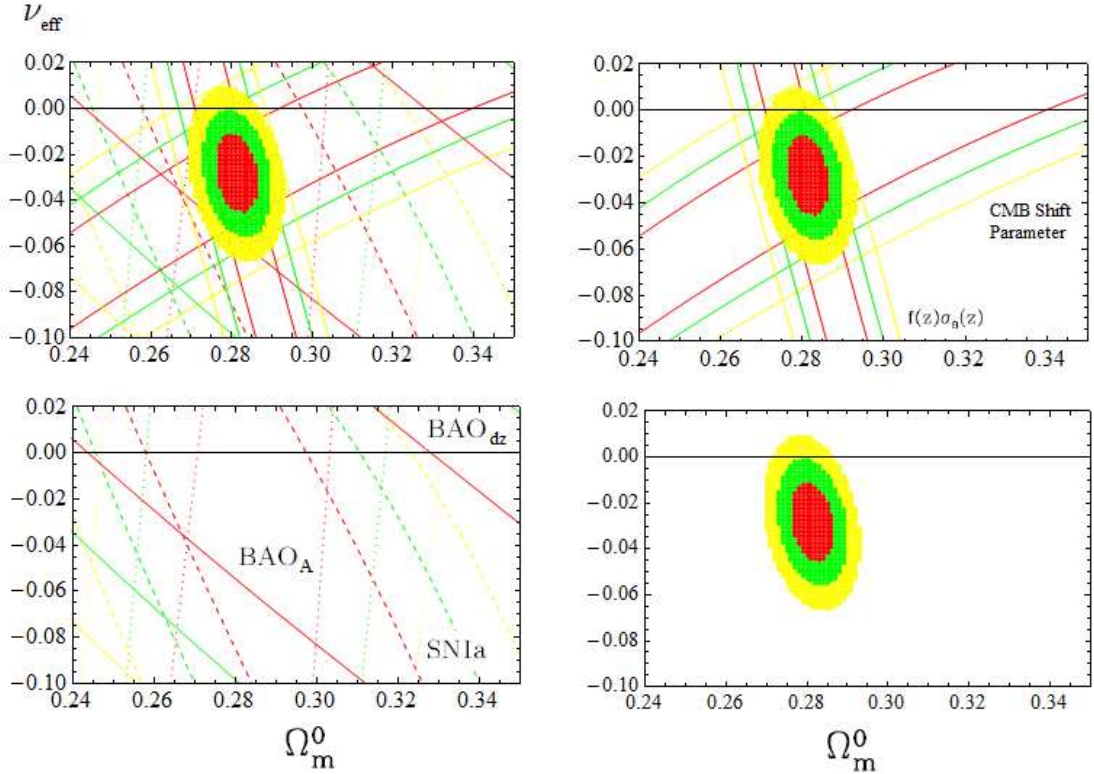}
\end{center}
\caption{Likelihood contours at $1,2,3\sigma$ C.L. in the $(\Omega_m^0, \nu)$-plane for the RVM (\ref{eq:ModelsA}) with $\alpha=0$ and $G=$const. using the expansion history (SNIa+BAO) and
LSS+CMB shift parameter data. In this scenario there is \emph{no} interaction with DM, implying that the DE is also self-conserved with a nontrivial dynamical EoS $\wD=\wD(z)$\,\cite{JCAP2015b}. The $\CC$CDM ($\nu=0$) appears excluded at $\sim 3\sigma$ level.}
\label{aba:fig1}
\end{figure}
In Fig. 1 we can see (on the left) the intersection of the SNIa, BAO and CMB shift parameter data for models (\ref{eq:ModelsA}) with $\alpha=0$, in which the vacuum ($\omega_D=-1$) is interacting with DM at fixed $G$. The BAO data used in this case makes use of the observations on the acoustic $A$-parameter defined by Eisenstein in \cite{Eisenstein05}, indicated as $BAO_A$. Similar contours can be obtained from the $BAO_{dz}$ data using the $d_z$-estimator, defined from the ratio of the acoustic horizon $r_s$ at decoupling and the dilation scale $D_V$ -- see \cite{JCAP2015a} for more details. In the panel on the right of Fig. 1 we display the corresponding likelihood contours at $1,2$ and $3\sigma$ confidence levels. As we can see, there appear hints of some preference of the data for $\nu\neq 0$ (the central value is at $\nu=+0.0048$, with $\Omega_m^0=0.282$) although the degree of evidence is at this point only moderate ($\sim 2\sigma$ C.L.).

In Fig. 2, we consider the same model (also under the assumption $\alpha=0$ and $G=$const.),  but now we assume local covariant matter conservation, which is expressed by the continuity equation $\dot{\rho}_m+3H(1+\omega_m)\rho_m=0$. The Bianchi identity enforces the conservation of the DE density, $\rD(H)$, and this implies that the EoS of the DE is not of the strict vacuum type ($\wD=-1$) but a dynamical function $\wD=\wD(H)$ (with $\wD(H_0)\simeq -1$) which satisfies the self-conservation law $\dot{\rho}_D+3H(1+\wD)\rho_D=0$. Furthermore, in this case we use the LSS data as a part of the total input data, i.e. we use BAO+SNIa+CMB plus the the linear growth data points, specifically the known data on the $f\sigma_8$ observable, see\,\cite{JCAP2015b} for details.  We observe from the likelihood contour lines in Fig. 2, obtained from the simultaneous combination of all these observables, that the physical region in the $(\Omega_m^0,\nu)$ plane is now more displaced from the $\CC$CDM line $\nu=0$ than in Fig. 1. In the current case and owing to the different set of hypotheses (self-conservation of matter and DE) the region that is selected in that plane is centered at a value $\nu<0$ (specifically $\nu=-0.028$), being five times larger in absolute value than in the case considered in Fig. 1. Amazingly enough, almost the entire $\sim 3\sigma$ region stays away from $\nu=0$. This means that the $\CC$CDM option with a rigid $\CC=$const.  is, in this case, excluded at a higher C.L. than before, essentially at $\sim3\sigma$ (equivalently at $99\%$ C.L.)!

\begin{figure}[t]
\begin{center}
\includegraphics[width=4in]{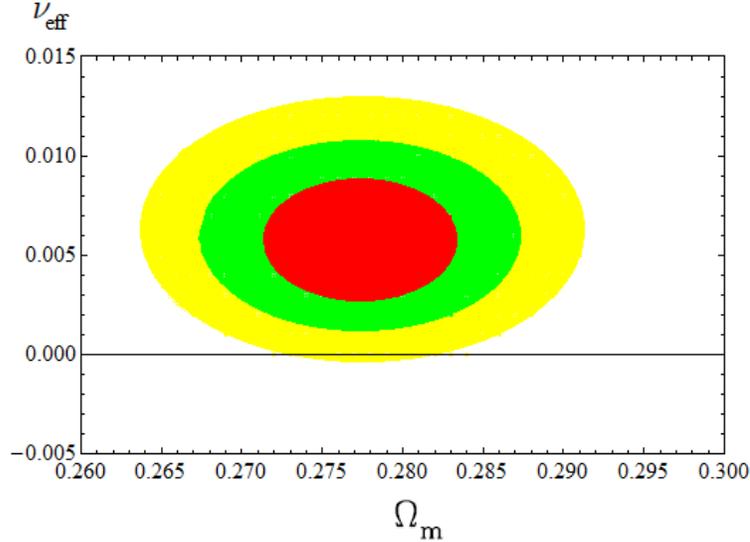}
\end{center}
\caption{Likelihood contours at $1,2,3\sigma$ C.L in the $(\Omega_m^0, \nueff)$-plane for the RVM (\ref{eq:ModelsA}) under matter conservation and running $G=G(H)$, using the full expansion history BAO+SNIa+H($z_i$) and LSS+CMB shift parameter data. The $\nueff=0$ region ($\CC$CDM) is $\sim 3\sigma$ away from the best value $\nueff=+0.0043$, and is therefore excluded at $99\%$ C.L. See Ref.\cite{ApJ2015} for details.}
\label{aba:fig1}
\end{figure}

Finally, in Fig. 3 we consider the general situation of the the RVM model (\ref{eq:ModelsA}) involving both $\nu$ and $\alpha$, and define $\nueff\equiv\nu-\alpha$. As before, we assume that matter (relativistic and nonrelativistic) is locally conserved, but now we treat $\rD(H)$ as vacuum energy (hence with the EoS $\wD=-1$). These conditions are compatible with the Bianchi identity if we assume that the gravitational coupling $G$ is slowly evolving. The precise conservation law reads $\dot{G}(\rma+\rr+\rL)+G\dot{\rho}_{\CC}=0$,  in differential form. This equation can be combined with (\ref{eq:ModelsA}) and Friedmann's equation, or equivalently using Eqs. (\ref{eq:generalizedFriedmann}), in order to integrate the model explicitly. The actual solution confirms that $G$ evolves very slowly. The exact formula is rather cumbersome, but it roughly yields $G(H)\sim \nueff\ln H$. Upon fitting the model to the overall SNIa+BAO+CMB+LSS data the contour plots of Fig. 3 can be derived\,\cite{ApJ2015}. It is rewarding to see that there is, once more, a marked preference of the data for a dynamical vacuum at an essentially $3\sigma$ C.L. (the central value lying at $\nueff=+0.0043$), which means that the $\CC$CDM model becomes anew excluded at $99\%$! Remarkably, the aforementioned $\sim 3\sigma$ C.L. deviation from the $\CC$CDM affords a significant improvement of the the joint likelihood fit to the cosmological data used; for details see\,\cite{ApJ2015} -- confer also the upcoming works\,\cite{Preparation2016} for an even more compelling evidence of running vacuum in the Universe (with or without interaction with DM). The advantageous and distinguished position of the RVM's is in stark contrast to other dynamical DE models, which are not able to improve the concordance $\CC$CDM, see e.g.\,\cite{LiZhangZhang15} and references therein.

%\subsection{You can safely ignore this}

%\section{Fitting Running Vacuum to Observations}

\section{Final remarks: Running Vacuum and the ``constants'' of Nature}

The detailed studies summarized here have shown that the idea that the cosmic vacuum should be dynamical in an expanding Universe is not only a theoretically appealing possibility but also a phenomenologically preferred option. The dynamics of the vacuum energy density (\ref{eq:ModelsA}) is effectively described in terms of the small parameter $\nueff=\nu-\alpha$, which plays the role of the coefficient of the $\beta$-function of the running $\rL$ and/or $G$.
The excellent current status of the RVM can be easily appreciated from the summary plots displayed mainly in Figs. 2-3, where the $\CC$CDM model is comparatively disfavored at $99\%$ C.L. The data are currently able to discriminate between the value $\nueff=0$ (corresponding to $\CC$CDM) and values of $|\nueff|\sim 10^{-3}$ at $3\sigma$ C.L. In actual fact the phenomenological situation of the RVM's is even better than indicated here. As it will be shown elsewhere\,\cite{Preparation2016}, the analysis of WMAP9, Planck 2013 and the recent Planck 2015 data, provides strong evidence that the  RVM class of dynamical models (\ref{eq:ModelsA}) is preferred as compared to the concordance $\CC$CDM. The precise meaning of ``strong evidence'' will be carefully quantified in terms of Akaike and Bayesian statistical criteria for model comparison\,\cite{Preparation2016}. It seems that the phenomenological support to the RVM, in detriment of the $\CC$CDM, has just begun.

Finally, let us mention that the RVM framework that we have outlined here has an additional bonus. It can also provide an explanation for the possible (slow) time-evolution of the fundamental constants of Nature\,\cite{ConstantsNature}. This is a field which probably holds many surprises in the future\,\cite{Preface}. The natural impact from the RVM on this issue occurs thanks to the cosmological exchange of energy between vacuum, matter and the possible interplay with the Newtonian coupling $G$. Because $\mu\sim H$ in Eq.\,(\ref{seriesRLH}), the RVM predicts that the associated rhythm of change of the fundamental constants (such as couplings, masses and vacuum energy density) should naturally be as moderate as dictated by the value of the Hubble rate at any given epoch. The RVM thus sets the natural time scale $1/H$ and a characteristic rhythm of variation of order $\dot{\cal P}/{\cal P}\sim H_0\,\Delta{\cal P}/{\cal P}\lesssim \left(\Delta{\cal P}/{\cal P}\right)\,10^{-10}$yr$^{-1}$ for any parameter ${\cal P}$, hence in the right ballpark. Typically $\Delta{\cal P}/{\cal P}\lesssim 10^{-3}$ over a cosmological span of time, which depends on the monitored parameter ${\cal P}=\CC, G, m_i,\alpha_{\rm em}, \alpha_s, \Lambda_{\rm QCD}$...\,\cite{FritzschSola}.

Such scenario intriguingly points to the possibility that there is a subtle crosstalk between the atomic world and the Universe in the large, which may be on the verge of being detected. We have called it elsewhere ``the micro and macro connection''\,\cite{FritzschSola2015}. It amounts to an almost imperceptible feedback between those two worlds and is responsible for a mild time drifting of the ``fundamental constants'' of Nature, in a way which is perfectly consistent with the general covariance of Einstein's equations\,\cite{FritzschSola}. Testing these ideas will soon be at reach of numerous experiments both at the lab (through a new generation of quantum optics experiments) and in the sky (through the astrophysical observations of molecular spectra in distant clouds); and, of course, also from the growing wealth of precision cosmological data. The incoming new era devoted to testing the micro and macro connection has just started. Most likely it will hint at the missing link between the physics of the very small and the physics of the very large\,\cite{Preface}, i.e. the (long sought-for) overarching interconnection of the subatomic quantum mechanical world with the large scale structure of the Universe -- and, ultimately, perhaps, the clue at solving the CC problem!

\section*{Acknowledgments}

I would like to take the opportunity to thank my senior collaborators S. Basilakos, H. Fritzsch, J.A.S. Lima and N. E. Mavromatos, and to my students  Javier de Cruz P\'erez, Adri\`a G\'omez-Valent, Elahe Karimkhani and Rafael Nunes, for discussions and/or collaboration on some of the materials presented here.
I have been supported in part
by FPA2013-46570 (MICINN), CSD2007-00042 (CPAN),
2014-SGR-104 (Generalitat de Catalunya) and MDM-2014-0369 (ICCUB).

\end{document}